\documentclass[aps,nature,british,floatfix,letterpaper,nonbalancelastpage,reprint]{revtex4-1}
\usepackage{graphicx} 
\usepackage{babel}
\usepackage{hyperref} 
\usepackage{amsmath} 
\usepackage{amssymb}
\newcommand{\ket}[1]{\left\lvert #1 \right\rangle}

\newcommand\startsupplement{%
    \makeatletter 
       \setcounter{table}{0}
       \renewcommand{\thetable}{S\arabic\c@table}
       \setcounter{figure}{0}
       \renewcommand{\thefigure}{S\@arabic\c@figure}
    \makeatother}

\usepackage{newfloat}
\DeclareFloatingEnvironment[name={Extended Data Figure}]{suppfigure}

\begin{document}
\title{Building logical qubits in a superconducting quantum computing system}

\author{Jay M. Gambetta}
\affiliation{IBM T.J. Watson Research Center, Yorktown Heights, NY 10598, USA}
\author{Jerry M. Chow}
\affiliation{IBM T.J. Watson Research Center, Yorktown Heights, NY 10598, USA}
\author{Matthias Steffen}
\affiliation{IBM T.J. Watson Research Center, Yorktown Heights, NY 10598, USA}
\date{\today}
\begin{abstract}The technological world is in the midst of a quantum computing and quantum information revolution. Since Richard Feynman's famous ``plenty of room at the bottom" lecture~\cite{Feynman:1960}, hinting at the notion of novel devices employing quantum mechanics, the quantum information community has taken gigantic strides in understanding the potential applications of a quantum computer and laid the foundational requirements for building one. We believe that the next significant step will be to demonstrate a quantum memory, in which a system of interacting qubits stores an encoded logical qubit state longer than the incorporated parts. Here, we describe the important route towards a logical memory with superconducting qubits, employing a rotated version of the surface code. The current status of technology with regards to interconnected superconducting-qubit networks will be described and near-term areas of focus to improve devices will be identified. Overall, the progress in this exciting field has been astounding, but we are at an important turning point where it will be critical to incorporate engineering solutions with quantum architectural considerations, laying the foundation towards scalable fault-tolerant quantum computers in the near future.
\end{abstract}

\maketitle

\section{introduction}

Quantum computing holds the promise of solving some computation problems that are untenable on conventional computers \cite{Shor:1997,Grover:1997,Yung:2014}. Loosely speaking, quantum computing targets problems that can exploit entanglement and superposition to explore a multitude of computational paths, then select the correct answer through constructive interference. For example, Shor's algorithm addresses the computational challenge of factoring by exploiting quantum interference to measure the periodicity of arithmetic objects~\cite{Shor:1997}.

However, there is a pernicious flaw to this increase in computational power. In a quantum computer, the information is encoded in quantum bits, or qubits, which need to interact strongly with one another, external inputs for control, and outputs for detection, but nothing else. This leads to the \emph{quantum conflict}: balancing just enough control and coupling, while preserving quantum coherence. This conflict represents a fundamental impediment to reducing the physical qubit error rate low enough to perform long/difficult/large-scale/practical quantum computations with them directly. 

Fortunately it has been shown that with quantum error correction it is possible to perform fault-tolerant quantum computing~\cite{Aharonov:1997,Preskill:1997}. The essential idea in quantum error correction (QEC) is to encode information in subsystems of a larger physical space that are immune to noise. QEC can be used to define fault-tolerant  \emph{logical qubits}, through employing a subtle redundancy in superpositions of entangled states and non-local measurements to extract entropy from the system without  learning the state of the individual physical qubits. The particular architecture for implementing a fault-tolerant operating scheme has bearing on the requirements necessary for the underlying physical qubits. 

While there are many approaches to achieving quantum fault-tolerance, one of the most promising is the two-dimensional (2D) surface code~\cite{Kitaev:2003,Bravyi:1998,Raussendorf:2007}. This code has a high tolerance to errors, or threshold (approximately $6.7\times10^{-3}$),  requires only nearest-neighbor qubit interactions, has simple error syndrome extraction circuits \cite{Dennis:2002}, and a suite of fault-tolerant logic based on transversal gates \cite{Bravyi:1998}, code deformation \cite{Raussendorf:2007,Raussendorf:2007a}, or lattice surgery \cite{Horsman:2012}.

All together, this suggests that to build a quantum computer we require:
\begin{itemize}
	\item	A physical qubit that is well isolated from the environment and is capable of being addressed and coupled to more than one extra qubit in a controllable manner,
	\item	A fault-tolerant architecture supporting reliable logical qubits, and
	\item	Universal gates, initialization, and measurement of logical qubits
\end{itemize}

A physical quantum computer satisfying all three of these requirements is still an outstanding challenge. However, in recent work, physical qubits in trapped-ion and superconducting systems have reached the point where errors are at or below the threshold~\cite{Chow:2012,Barends:2014,Harty:2014aa, Nigg:2014}, and networks of 4-9 superconducting qubits with individual control and readout have been used to show concepts of error correction \cite{Corcoles:2015,Riste:2015,Kelly:2015}. Over the next few years the field will be at a stage of building interesting quantum devices with a complexity that could never be emulated in full generality on a classical computer ($\sim 50$ or more qubits). These devices will allow us to understand nature in a regime that has never been explored before,  offering new insight into analog simulations of quantum systems. Nonetheless, near term progress towards the monumental task of fully fault-tolerant universal quantum computing will hinge upon using QEC for demonstrating a quantum memory: a logical qubit that is sufficiently stable against local errors and ultimately allows essentially error-free storage.  

In this Review we present a view of what a medium-sized quantum computing system will comprise as well as a discussion on the current state of coherence, control and readout of superconducting qubits. We also detail some of the main challenges which the community will face to build a device of O(100) qubits so as to further the state-of-the-art of QEC and approach a useful fault-tolerant quantum memory.

\section{The quantum computing system and fault-tolerant architecture}

\begin{figure}
	\centering
	\includegraphics[width=3.375in]{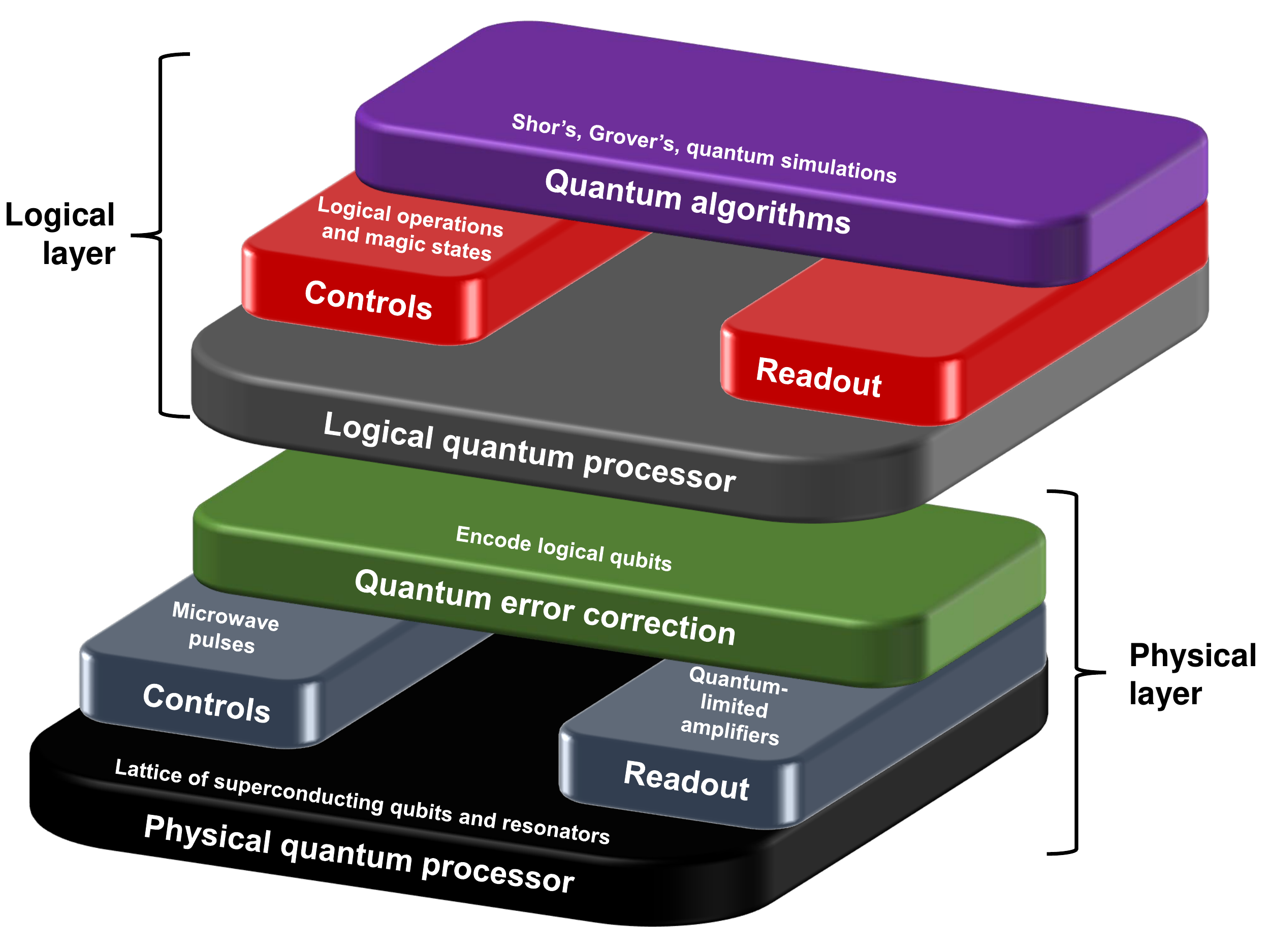}
	\caption{\label{fig1} A systems view of a quantum information processor. It consists of a physical and logical layer. The physical layer provides the error correction and consists of a physical quantum processor that has both input and output lines that are controlled by the quantum error correction processor. This processor is in turn controlled by the logical layer where the encoded qubits are defined and the logical operations are performed for the desired quantum algorithm.}
\end{figure}

A full quantum computing system can be envisioned within a layered structure as shown in Fig.~\ref{fig1}. The system is comprised of two primary layers, a physical qubit layer, and a logical qubit layer. The lower physical layer contains physical qubits controlled via a QEC processor, which is in essence a classical processor that uses measurement outcomes of the physical qubits to realize a QEC code. This classical processor keeps track of the physical errors which arise, and implements the appropriate feedback on the controls of the physical qubits. The upper layer of Fig.~\ref{fig1} is called a logical layer and functions through control of the physical layer. Here, logical qubits are encoded within the fully error-corrected system of physical qubits, and logical controls and readouts are governed through a processor that determines how to implement difficult quantum algorithms, e.g. Shor's, Grover's, and quantum simulation \cite{Shor:1997,Grover:1997,Yung:2014}. 

Residing at the very bottom of the physical qubit layer are imperfect physical qubits, which have been explored in a variety of different experimental systems: superconducting qubits~\cite{Devoret:2013,Corcoles:2015,Riste:2015,Kelly:2015}, trapped-ion~\cite{Ospelkaus:2011aa,Nigg:2014,Harty:2014aa}, solid-state spin~\cite{Shulman:2012,Zwanenburg:2013aa}, nuclear spin~\cite{Gumann:2014,Lu:2015}, non-linear photonic~\cite{Martin-Lopez:2012aa}, and neutral atom~\cite{Maller:2015aa} qubits, which are just a few examples of more mature qubit systems  demonstrating multi-qubit operations. Coherence times in these systems are relatively varied. However, for the purposes of quantum computing, it is important to normalize coherence to the gate (control) lengths possible for the system. Especially for superconducting qubits, with coherence times in the $\sim$$100$~$\mu$s range and gate lengths $\sim$10-100 ns, the number of operations per coherence time becomes a very promising number (currently  approaching $10^4$ operations).  

\begin{figure}
	\centering
	\includegraphics[width=0.7\linewidth]{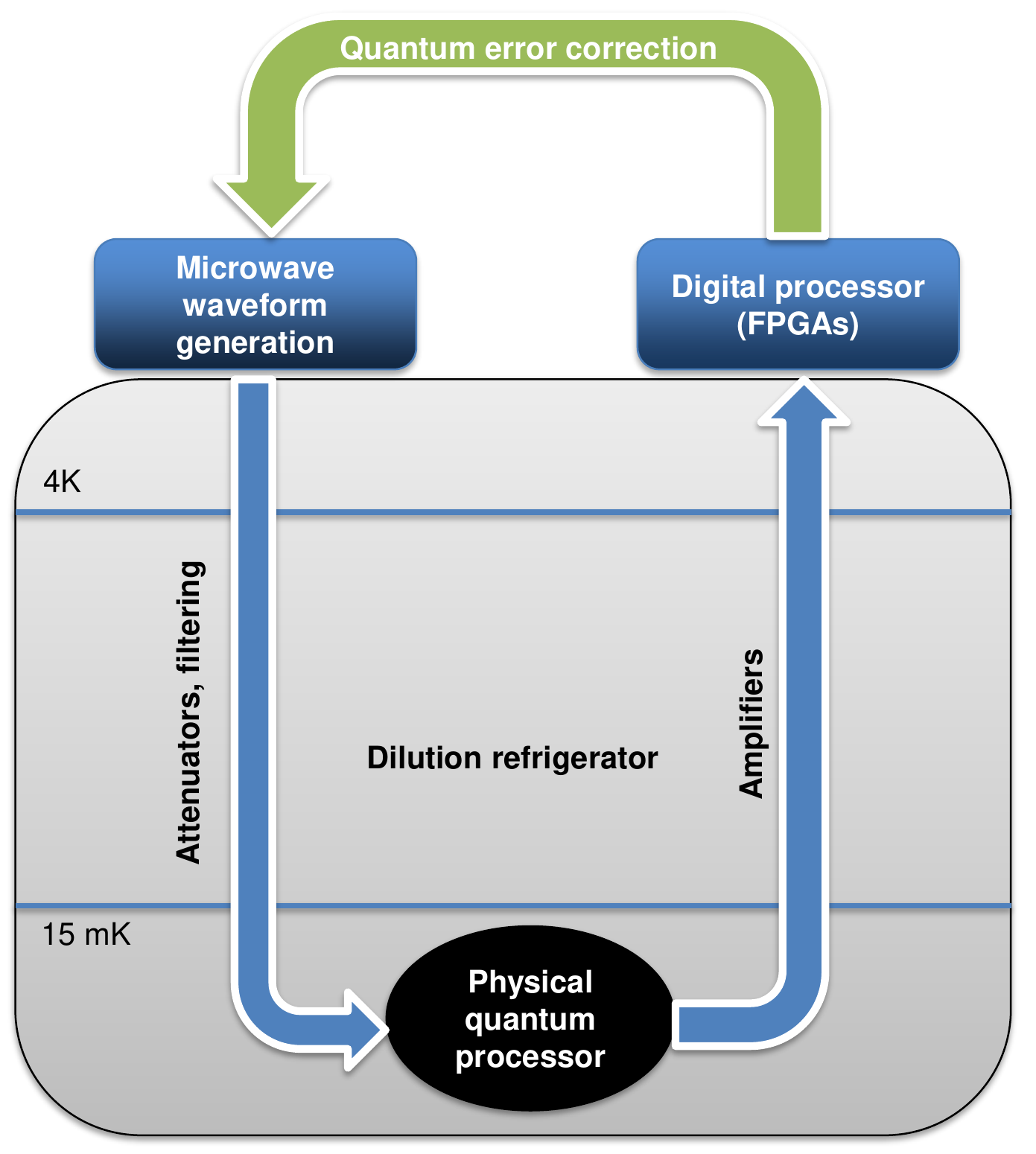}
	\caption{\label{fig:figure_schematic} The physical qubit processor is located at the bottom (15 mK) plate of a dilution refrigerator. Microwave pulses are generated at room temperature using synthesizers, arbitrary waveform generators, and mixers. These pulses are filtered and attenuated to assure negligible noise at the qubit. High-fidelity readout of the qubit state requires quantum limited and other cryogenic amplification to overcome thermal noise for digitization and weighted homodyne measurement\cite{Magesan:2015}. The quantum error correction processor sits above and orchestrates the physical control and readout functions, to perform the error correction protocol.}
\end{figure}

Inputs and outputs to the layer of physical qubits are controls and readouts, respectively. For superconducting qubits, controls can involve a full suite of microwave electronics and pulse-shaping, to realize specific qubit rotations and two-qubit controlled operations. The noise on these controls is filtered and attenuated so that the noise at the qubit is negligible.  It is also important for readout of superconducting qubits to be boosted through stages of amplification, the first of which is quantum-limited \cite{Hatridge:2011aa,Abdo:2011aa,Hover:2014aa}. Analog readout signals are then digitally processed either on classical computers or in customized field-programmable gate arrays (FPGAs) for fast processing~\cite{Riste:2012aa,Vijay:2012aa}.  A schematic of the physical layer for superconducting qubits is shown in Fig.~\ref{fig:figure_schematic}. Although currently the quantum error correcting process sits at room temperature, it is possible in the future that some of this operation might be performed at lower temperatures stages within the dilution refrigerator~\cite{Hornibrook:2015aa}. 

\begin{figure*}
	\centering
	\includegraphics{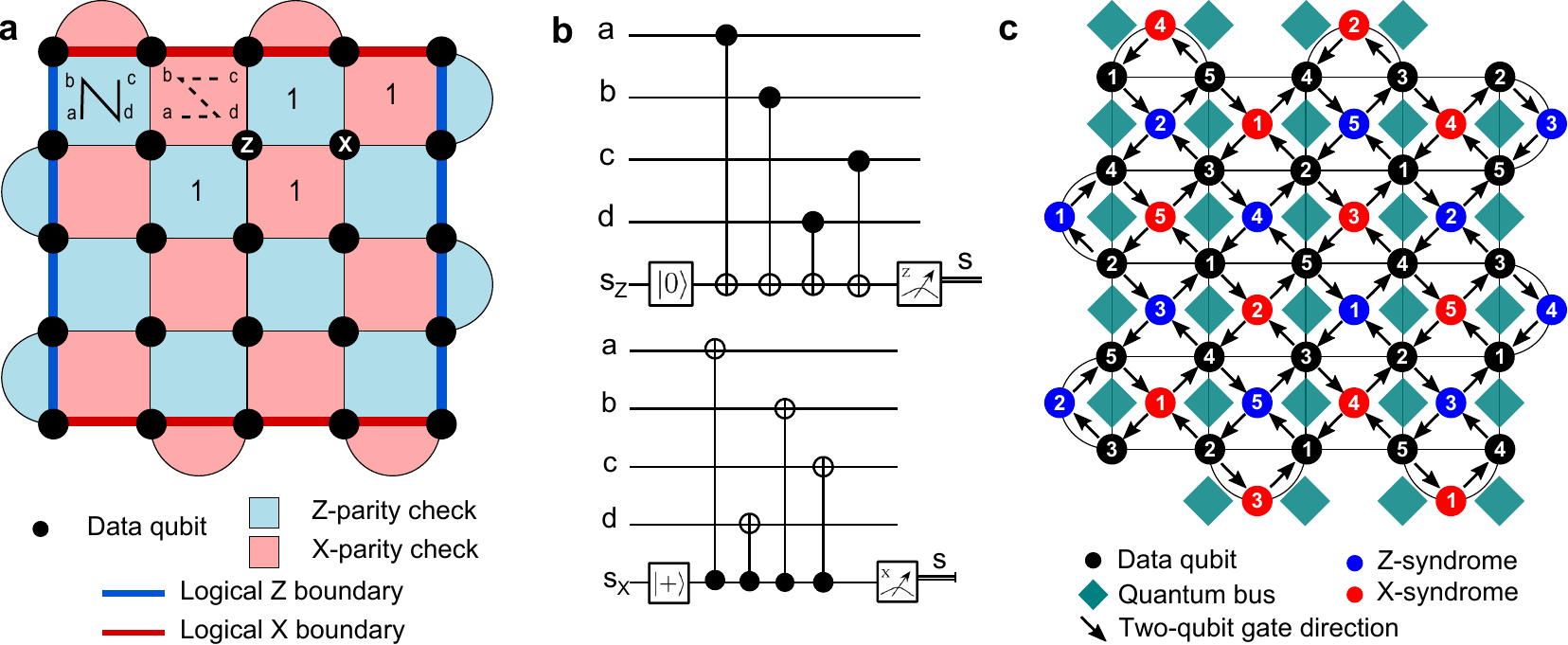}
	\caption{\label{fig:Fig2}(a) The rotated surface code (RSC) with $d^2$ data qubits (black circles) and the $(d^2-1)$ $X$- (red) and $Z$-parity (blue) checks, shown for $d$=5. Examples of errors are shown (white lettering inside black data qubit circles) along with the corresponding measured syndromes for surrounding parity checks. Logical operation boundaries are shown for $X$ and $Z$ Pauli operations as red and blue lines, respectively. (b) The quantum circuits used to perform the $Z$-parity and $X$-parity checks. The order of these checks ensures that no hook errors \cite{Dennis:2002} can be incorrectly assigned as single-qubit errors. (c) A physical realization of the RSC superconducting qubits with quantum buses for mediating interactions. Here, quantum buses are used to connect data qubits to syndrome bits for $Z$ and $X$ parity detection. Furthermore, a proposed arrangement of five distinct qubit frequencies (numbering within the qubits) is shown to minimize crosstalk and addressability errors~\cite{Gambetta:2015}.}
\end{figure*}

The particular arrangement of physical qubits is governed by selection of a fault-tolerant error correction architecture. With high error thresholds and simple physical lattice arrangement, the rotated surface code (RSC) \cite{Bombin:2007} is one of the most promising schemes for QEC. Fig.~\ref{fig:Fig2}\textbf{a} shows a conceptualized version of the RSC consisting of $d^2$ data qubits (the black dots at the vertices of the graph) arranged in a square lattice where $d$ is the distance of the code (shown for a $d$=5 code). These qubits are in a simultaneous eigenstate of both the ($d^2-1$) $Z$-parity (blue faces) and $X$-parity (red faces) stabilizers and the logical qubit is the one remaining unconstrained degree of freedom. The $Z$($X$)-parity stabilizers are the multi-qubit operators represented by the product of single qubit $Z$ ($X$) operator on the data qubits located at the vertices of the face. A logical $\bar{X}$ (Pauli $X$ bit-flip logical gate) is represented by a chain of $X$ operations that connect the two $X$-boundaries (red lines) and a logical $\bar{Z}$ (Pauli $Z$ phase-flip logical gate) is represented by a chain of $Z$ operations that connect the two $Z$-boundaries (blue lines). 

The QEC processor (green slab in Fig.~\ref{fig1}) sits above the physical controls and readouts and functions to keep the lattice in a simultaneous eigenstate of the $Z$-parity and $X$-parity stabilizers implemented by the circuits shown in Fig.~\ref{fig:Fig2}\textbf{b}. The order of the controlled-NOT (cNOT) gates is important for fault-tolerant operation to ensure that all error syndrome bits correctly identify single error faults anywhere in the extraction process.  In each cycle, $d^2-1$ syndrome bits are extracted, and in the absence of errors the syndrome bit will have the same value every cycle. Each time the syndrome bit changes value, an endpoint of a chain of errors is identified. From these chains, Edmonds'€™ minimum weight perfect matching algorithm can be used to find the set of corrective operations~\cite{Edmonds:1965}. Corrections are applied to the classical data associated with the measurement results rather than to the actual physical qubits.  This ensures that no corrective operations need to be applied to the qubits (no additional errors) and no complicated feedback is necessary. 

\begin{figure}
\centering
\includegraphics[width=\linewidth]{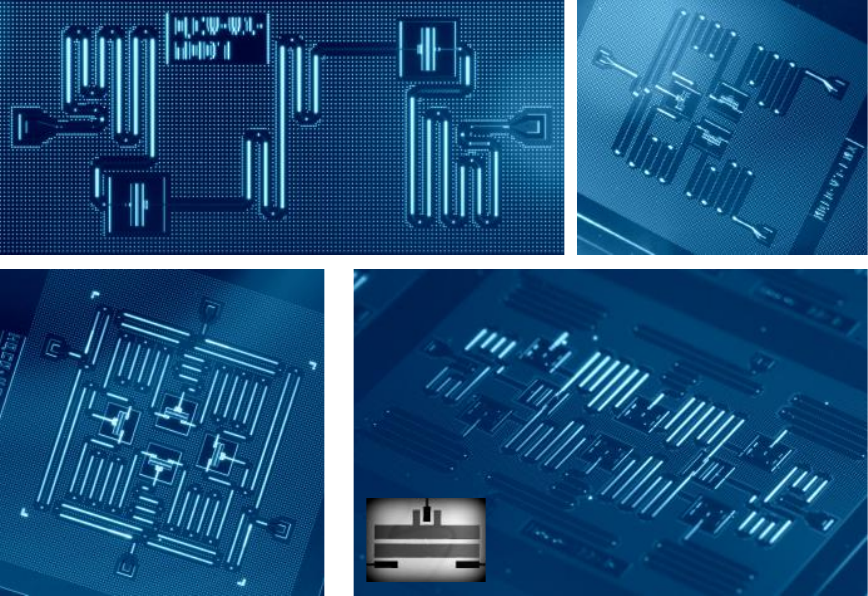}
\caption{\label{fig:Fig3} Images of four recent devices fabricated at IBM.  The device in the top left corner contains 2Q(ubits)/1B(us)/2R(eadout resonators) and is currently being used to study optimal two-qubit gates. The top right corner shows a device with 3Q/2B/3R which was used to demonstrate a parity measurement~\cite{Chow:2014}. In the lower left corner, is a device with 4Q/4B/4R for demonstrating the [[2,0,2]] code~\cite{Corcoles:2015} and the lower right corner shows a device of 8Q/4B/8R for studying both $Z$ and $X$ parity checks. Inset shows an optical micrograph of an individual transmon qubit. }
\end{figure}

To realize this code we induce coupling between the data qubits and the syndromes by using a quantum bus~\cite{Majer:2007,Sillanpaa:2007}. The arrangement of the bus is shown in Fig.~\ref{fig:Fig2}\textbf{c} as the green squares, where each bus couples to four data qubits and each qubit couples to two buses allowing a tiling that achieves the connectivity required for the RSC \cite{Chow:2014}. Using this tiling and provided the gate is directional (e.g. cross-resonance \cite{Chow:2011} and tunable-frequency activated gates\cite{DiCarlo:2009}) a minimum of five frequencies are required to allow selective two-qubit gates~\cite{Gambetta:2015}. At IBM we are using coplanar waveguide microwave resonators for the bus and examples of connected multi-qubit experimental devices employing this sub-lattice are shown in Fig.~\ref{fig:Fig3}.

With an error-corrected sea of physical qubits, it becomes possible to enter the logical layer of Fig. \ref{fig1}. In the RSC only the logical Hadamard can be implemented transversally (i.e. the logical gate is represents by the product of operations on the single qubits) up to a rotation of the code. The version of the cNOT that we find  attractive because of its lower overhead is provided by lattice surgery \cite{Horsman:2012}.  Optimal implementations of the remaining logical gates ${S}=\sqrt{Z}$, ${T}=\sqrt{S}$ are still open questions. It is known that RSC or any 2D stabilizer code can not implement all universal gates transversally \cite{Bravyi:2013} and methods such as distillation and injection are needed \cite{Bravyi:2005}.  The distillation process for the $T$ gate is the most expensive in terms of number of physical qubits and the community is working hard to reduced this overhead \cite{Bravyi:2015,OConnor:2015}.  With these logical operations all quantum gates can be implemented efficiently using the Solovay-Kitaev algorithm \cite{Dawson:2005} or more recent optimizations \cite{Kliuchnikov:2013,Selinger:2015}. At the top of this layer sits the proposed applications of quantum computing, including the possibility to perform quantum algorithms such as Shor's factoring \cite{Shor:1997}, Grover's search \cite{Grover:1997}, or digital quantum simulations of real world chemical molecules and dynamics\cite{Yung:2014}.

For the rest of this article, we focus primarily on the lower section of the diagram, giving some detail about the current state of such a quantum computing system with respect to superconducting qubit technology implementing the RSC approach for quantum error correction. We will discuss some of the challenges that lie ahead for the superconducting qubit community in the march towards implementing a fully-error corrected surface of physical qubits. 

\section{Coherence of Superconducting qubits}

Josephson junction (JJ) based superconducting qubits have emerged as one of the contenders to build a practical quantum computer. Detailed descriptions of superconducting qubits have been discussed previously~\cite{Clarke:2008aa,Devoret:2013} and here we only provide a brief historical overview and concise description of current state-of-the-art implementations.

Superconducting qubits are constructed out of one or more inductors, capacitors and JJs. By virtue of employing superconducting materials, losses in these systems can be greatly reduced which is a necessary requirement. The JJ consists of a superconducting electrode, a thin insulator and another superconducting electrode allowing Cooper pairs to tunnel across the insulating barrier and behaves effectively as a non-linear inductor. This circuit element is necessary because its non-linearity leads to quantum mechanical energy levels that are not equally spaced. The deviation from a harmonic oscillator (or anharmonicity) permits the isolation of two energy levels to provide the computational basis states $\ket{0}$ and $\ket{1}$ for a qubit. Typically, qubit transition energies (between $\ket{0}$ and $\ket{1}$ are in the microwave frequency regime, $\sim4-6$ GHz,  and the difference between this transition and one that involves going out of the qubit manifold ${\ket{0},\ket{1}}$ is $5 \%$  for weakly anharmonic qubits like the transmon qubits \cite{Koch:2007aa}.

Important metrics for the qubits are the coherence times, $T_1$ and $T_2$ (energy relaxation and dephasing times, respectively). Energy relaxation quantifies the time it takes for a qubit to decay from its excited state $\ket{1}$ to the ground state $\ket{0}$ (a bit-flip error) while dephasing times correspond to the time it takes for a quantum superposition state $\ket{+}=(\ket{0}+\ket{1})/\sqrt{2}$ to lose its phase relationship between $\ket{0}$ and $\ket{1}$ (i.e. a phase-flip error). Both quantities play an important role as shorter times will reduce the accuracy of quantum operations.

The first experimental demonstration of a superconducting qubit~\cite{Nakamura:1999aa} is attributed to the group at NEC in 1999 albeit with $T_2 \sim 1$ ns. Since this seminal result, many groups around the world conceived and implemented a variety of superconducting qubits by varying the superconducting circuits, for example, by adding loops interrupted by one or more JJs or by adding capacitors. Research involving all these variants helped the community shed light on what limits coherence times. By now it is known that charge noise, flux noise, the microwave environment, and materials play crucial roles.

Any electromagnetic mode with finite quality factor that  couples to the qubit will impact the  $T_1$. $T_1$ limitations from various external couplings can now be analytically calculated, by analyzing the real part of the admittance as seen by the qubit~\cite{Houck:2008aa}. As such, over the years many results have shown how to reduce or eliminate residual coupling to electromagnetic modes that are present, intentional or not~\cite{Reed:2010,Jeffrey:2014aa,Bronn:2015}. At the same time it is also necessary to minimize stray radiation especially at high frequency which may be capable of generating quasi-particles~\cite{Barends:2011,Corcoles:2011}.

Dielectric loss plays a crucial role and appears to be limiting $T_1$ for many superconducting qubits. It is believed to be due to two-level systems (TLSs) at the microscopic level~\cite{Martinis:2005} that couple to the qubit's electric field~\cite{Wenner:2011,Wang:2015,Dial:2015}. This dielectric loss manifests itself in two different ways. First, bulk insulating material with a non-zero loss tangent that is involved in any of the qubit's total capacitance can limit $T_1$ times. A continuum of TLSs residing in this bulk material lead to the standard exponential decay. However, when there are only few TLS present the dielectric loss manifests itself differently. Individual TLSs at some specific frequencies can couple to the qubit and give rise to avoided level crossings among other undesirable effects~\cite{Martinis:2005}.

For the transmon qubits which we design~\cite{Chow:2015}, we typically aim for a transition frequency of 5-5.4 GHz, with an anharmonicity of $\sim-346$ MHz so that the charge dispersion is less then 30 KHz. With numerical simulations combined with static field simulations of the qubit design, we aim to to achieve a qubit capacitance of $C_\mathrm{q} \sim 65$ fF, paired with a JJ critical current of about $I_0 \sim 27$ nA. The junction is made quite small ($100-200$ $\mu\text{m}$  $\times$ $100-200$ $\mu\text{m}$ ) which avoids TLS defects residing in the tunnel junction. The shunting capacitor is formed by metal pads spaced apart as much as 70 $\mu$m so as to minimize dielectric loss from any of the substrate surfaces~\cite{Wang:2015,Dial:2015}. This style of qubit currently provides some of the highest and reliable coherence times for transmon devices,  $T_1,T_2  \sim 100 \mu$s, almost 5 orders of magnitude improved over the initial demonstration of superconducting qubits and enough to demonstrate concepts of error correction.

\section{Control of Superconducting Qubits}

To universally control  a quantum system, it is sufficient to be able to perform arbitrary single-qubit gates and a two-qubit gate \cite{Barenco:1995}. For superconducting qubits most researchers have converged on using microwave drives to perform arbitrary  single-qubit rotations in the $x-y$ plane through control of the amplitude and the phase of the drives. However, for transmon qubits, due to the weak anharmonicity, it is necessary to perform corrections due to the effects of the higher levels. The standard is to use Derivative Removal Adiabatic Gate (DRAG) shaping \cite{Motzoi:2009}. This approach has improved single-qubit gate fidelity to $5(2)\times10^{-4}$ as demonstrated by benchmarking\cite{Sheldon:2015}. Interestingly, coherence times predict that these gates should still be much better, and it is still an open question as to what is the limitation \cite{Sheldon:2015}. 

There have been many variants of entangling two-qubit gates for superconducting qubits, each with their own set of pros and cons. We find it convenient to split the gates into two classes. One class of gates contains all of those which rely on the dynamical flux-tunability of either the underlying qubits, or some separate sub-circuit. This includes the direct-resonant iSWAP  \cite{Bialczak:2010,Dewes:2012}, and the higher-level resonance induced dynamical c-Phase (DP) \cite{DiCarlo:2009,Barends:2014}. The second class of gates contains all those in which the qubits have fixed-frequencies and microwave sources are used to activate the interaction. The gates in this class include the resonator sideband induced iSWAP \cite{Leek:2009}, the cross-resonance (CR) gate \cite{Chow:2012}, the Bell-Rabi \cite{Poletto:2012}, the mircowave activated phase gate \cite{Chow:2013}, and the driven resonator induced c-Phase (RIP) \cite{Cross:2015}.

The primary advantage of the dynamically tunable class of gates is the ability to operate the qubits in very different regimes: one in which the qubits are independent with negligible interaction, and one where the two-qubit interaction is maximized. In the first regime, single-qubit gates can be applied trivially without the need for specialized decoupling schemes as the qubits will not experience significant crosstalk errors. In the second regime, the qubits can be tuned to optimize the two-qubit interaction so as to enable the shortest possible gate times. This allows simple single-qubit gates, the possibility of strong two-qubit interactions, and low crosstalk errors. The main disadvantages of such gates are the reliance on flux-tunable qubits, which can have reduced coherence times due to flux-noise \cite{Yoshihara:2006}, the risk of interacting with other energy levels in the system during tuning, and additional circuit and control complexity due to on-chip tunable flux controls or couplers which support dynamical tunability.

For the case of the microwave two-qubit gates, the qubits are fixed in frequency, and thereby can be parked at `sweet-spots' of coherence or made to be untunable. Furthermore, the addressing hardware and shaped-microwave controls become analogous to those of single-qubit gates. There is additional circuit complexity for some of the schemes, specifically CR gates require local microwave addressability for each qubit. The most significant disadvantages for the fixed-frequency gates are tradeoffs to either coherence or single-qubit control in order to have stronger two-qubit interactions. For example, in the CR gate the qubit-qubit detunings which would give the strongest two-qubit interaction, also happen to result in reduced single-qubit addressability. This potentially could be overcome by optimal control. 

,
Fast high-fidelity two-qubit gates is still an open area of research, although great progress has been made with the DP gate\cite{Barends:2014} achieving $6\times10^{-3}$ and we have achieved $1.4(2)\times10^{-2}$ with the CR gate and $2.3\times10^{-2}$ with RIP. Moving forward these gates need to be made better and a good goal for the field is to achieve errors below $10^{-4}$, $\sim 100$ times less than the RSC threshold so that the overhead is not too large. As these fidelities continue to improve, important questions emerge in verification and validation. It becomes critical to bound the errors for different characterization methods: for example although randomized benchmarking \cite{Knill:2008,Magesan:2011} has become a standard for the field, it is also known that this method fails to pick out particular types of errors \cite{Epstein:2014}. 

\section{Readout of Superconducting Qubits}

To discern the state of a superconducting qubit, it has become standard to use a dispersive interaction with a resonator \cite{Blais:2004,Wallraff:2004,Gambetta:2008}.  This interaction results in a dispersive shift that causes the frequency of the resonator to change depending on the state of the qubit \cite{Gambetta:2008}. The resonator frequency is
interrogated with a microwave pulse, typically at a frequency near the midpoint of the resonant frequencies corresponding to the ground and excited states, and the phase and amplitude of the reflected signal are used to distinguish the state of the qubit. The appropriate discrimination techniques (including machine learning \cite{Magesan:2015})  and optimal control protocols for depopulation and resetting of the cavity states~\cite{McClure:2015} are areas of important exploration for realizing repetitive error correction protocols required for syndrome detection.

However, a large part of the focus for improving readout is dealing with the circuitry after the resonator, including elements such as filters and amplifiers. Purcell filters~\cite{Reed:2010,Jeffrey:2014aa,Bronn:2015} are commonly used to reduce the probability of the qubit to undergo spontaneous emission (and hence excited state relaxation) by changing the coupled external environment at the frequency of the qubit. Although original incarnations~\cite{Reed:2010a} had a very small frequency bandwidth of protection, since then better designs have allowed for larger bandwidth versions~\cite{Jeffrey:2014aa,Bronn:2015} which should allow for $T_1$ to be in principle well over 10 ms, while still having enough signal for high-fidelity readout.

Quantum limited amplifiers are used to further boost the readout resonator signals, and a number of parametrically-driven options are now widely available. These include Josephson parametric amplifiers, which incorporate a microwave resonator with an inherent non-linearity, parametrically driven so as to activate a three-wave-mixing effect~\cite{Hatridge:2011aa}. Other amplifiers also incorporate four-wave mixing via a Josephson-ring modulator, with the added benefits of frequency conversion or non-reciprocity~\cite{Bergeal:2010}. Besides these amplifiers which add the minimal half-photon of quantum noise, near quantum-limited amplification with demonstrated directionality are also being investigated using traveling-wave effects and SQUIDs~\cite{Ho-Eom:2012,Hover:2014aa,OBrien:2014}. In total, the level of sophistication for qubit-readout amplifiers has matured significantly such that high-fidelity single-shot readout is a well-known experimental technique. 

\section{Technical challenges ahead}

The technical advances over the last 15 years for superconducting qubits have been astounding. However, it is important to ask what is now necessary for a demonstration of quantum computing on a modest sized system and what such a demonstration might look like. With current experiments scaling into double-digit number of qubits, a lattice of O(100) physical qubits which can perform quantum error correction experiments is well within possibility in the near term. Such a system would serve as an invaluable learning tool not just for testing the feasibility of QEC, but also for enabling insight into how to scale a system to the next level of $10^4$ to $10^8$ physical qubits. With such numbers of physical qubits, some of the canonical quantum algorithms could possibly be tested in a universal fault-tolerant system. Getting to this important intermediary stage of O(100) qubits would represent a major stepping stone towards bringing the next level of quantum computing to reality.

There are still a number of important technological challenges to address to successfully demonstrate an O(100) qubit system. Aside from advances in coherence times, optimal control and calibration routines for high-fidelity quantum gates, the following list represents other critical areas for exploration and advancement: 
\begin{itemize}
	
\item \emph{Breaking the plane.} Arrangements of multi-qubit devices to-date have been limited to a single physical plane.  This has serious limitations for systems beyond an $n\times2$ square lattice. In such a scenario, qubits on the interior of the grid, require a path in and out for addressability. There are many options for breaking this ``third dimension" which include standard silicon-based lithographic techniques such as thru-silicon-vias, a flip-chip multi layer stack, or employing waveguide package resonance modes~\cite{Minev:2015}. Ultimately, the solution must also be  cryogenically compatible, preserve coherence times and gate fidelities, while not also introducing any new loss mechanisms. 
	
\item \emph{Substrate modes.} The device substrate size will need to increase in order to accommodate a larger number of qubits. Due to the boundary conditions of the substrate die, it will host electromagnetic modes that decrease in frequency for increasing sizes. These modes can facilitate both cross-talk between pairs of qubits in the plane or a reduction in coherence times. While at the moment this problem can be circumvented by clever design, this challenge in the long term is an outstanding question. Metallic vias are a potential route, although these must be cryogenically compatible and the additional fabrication processing and materials must also not negatively impact coherence times.

\item \emph{On-Chip Microwave Integrity.} As the complexity of the network grows it becomes more difficult to insure that  broken ground planes in the designed network are still properly tied together at relevant microwave frequencies. Improper grounding can result in undesired slot-line modes and other spurious microwave resonances which again will lead to cross-talk and reduced coherence. Air-bridge cross-overs ~\cite{Chen:2014}, vias, and flip-chip lids~\cite{Abraham:2014} are all potential paths towards improved microwave integrity.
	
\item \emph{Josephson-Junction Reproducibility \& Accuracy.} In a large network of qubits, variations in the qubit frequency will lead to undesired frequency collisions of the fundamental and higher levels. Such collisions can lead to strong correlated interactions, leakage effects, and addressability errors. Currently, numerical simulations of qubit device designs have allowed an accurate prediction of the qubit capacitance and the couplings. However, fluctuations in the critical current of the Josephson junction are currently on the order of $10\%$, which result in a $\sim$280 MHz variation in our designed qubits. This is a substantial spread which will only be improved through more reliable Josephson junction fabrication. Another aspect is the long-term critical current variability of fabricated Josephson junctions, and investigating what might influence perturbations on successive experimental cooldowns.

\item \emph{Extensible Control and Readout Hardware.} With current qubit network devices, the usage of commercial-off-the-shelf (COTS) equipment for control and readout is not yet cost prohibitive. However, moving towards O(100) networks would require a substantial lowering of the cost per qubit. This can be achieved by shifting from COTS towards customized and targeted electronics solutions. Nonetheless, another important caveat to consider is the overall noise performance for different hardware (e.g. phase, amplitude noise), and ensuring that these would not limit ever decreasing gate and readout error rates. The extensibility of readout hardware for the application towards QEC also hinges upon having low latency, and the ability to perform a full qubit state discrimination, meter and qubit reset, at a fast desired measurement rate. This could potentially involve customized design for fast-feedback on FPGAs, which are also amenable to programming new concepts for discrimination~\cite{Magesan:2015}.

\item \emph{Cryogenic System Integrity.} Larger devices also mean more signal-carrying wires and ancillary microwave equipment which all sit inside of a dilution refrigerator. The cryogenic load will need to be handled with care, especially in the proper appropriation of filtering, attenuation, isolation, and amplification so as to not degrade any coherence times, while providing the ability to perform fast and high-fidelity operations. The exact engineering of the cryogenic system environment (e.g. thermalization, impedance matching, infrared radiation shielding~\cite{Corcoles:2011, Barends:2011, Rigetti:2012}) and paths towards reduction of  component size and mass for isolators, amplifiers, circulators, etc., are important topics of open study.

\item \emph{System Calibration.} State-of-the-art  high-fidelity gate experiments~\cite{Sheldon:2015,Kelly:2014} have already shown that the accuracy of gates can crucially depend on the ability to calibrate all necessary microwave pulse parameters. How, then, will the complexity of the calibration set grow with system size?
 With more connected qubits, the possibility of correlated errors increase, and new sequences need to be developed to ensure that these play no significant effect on the performance of the independent controls. Determining how to make robust and extensible system calibrations will be critical for high-performance experiments.

\item \emph{Verification \& Validation.} Tools currently exist to measure the accuracy of one or two-qubit gates in a relatively straight forward manner. These can in principle be extended to larger systems but typically scale exponentially with increasing number of qubits \cite{Chuang:1997} or give only partial information \cite{Knill:2008,Magesan:2011,Magesan:2012a}. Moving forward, tools that determine how accurate quantum gates operate on a subset of a larger fabric of qubits will have  relevance for QEC. One current technique, simultaneous benchmarking \cite{Gambetta:2012}, is a starting point, but it is not yet clear how sensitive it can be towards adverse errors.  Overall, verification and validation methods will likely grow from bootstrapping techniques on smaller subsystems, extended upwards towards larger lattices.

\item \emph{New QEC Codes}. Even though the surface code and its variants are very attractive for guiding current experiments, there is still a significantly large overhead associated with proper functioning, especially when going towards logical qubit operations. The community is actively working towards reducing this overhead through either finding new codes with inherent universal transversal operations~\cite{Bravyi:2015,OConnor:2015} or by  reducing  the requirements for magic state distillation~\cite{Bravyi:2012aa,Jones:2013}. 
	
\item \emph{Software.} In the end, when an O(100) system is built, how would such a system be operated? It still remains to be seen what will be the assembly language of such a quantum network, to take simple instruction sets in a user-defined program, to be translated into sequences of microwave and/or flux signals that are applied to the qubits directly. Moving forward, a standardized software and compiler code-base will need to be developed. 
	
\end{itemize}

\section{Conclusion}

In conclusion, the road towards a full-scale universal quantum computer with fault-tolerant error correction will be long, dark, but filled with exciting challenges. However, in the near future, systems of O(100) qubits are within reach and already beyond what can be emulated in full generality on a classical computer. This will usher in a new era in quantum information science, with explicit hardware to match broad ideas in theory, and culminate in the  demonstration of a useful logical memory. A very promising route forward is the rotated surface code, offering simplicity in the network, implemented with superconducting qubits,  which has shown tremendous and rapid progress in coherence times, controls, and readout.  The challenges which we outlined in this review while difficult are not insurmountable and with clever engineering and new insights we believe the road ahead is lit.

\begin{acknowledgments}
The authors would like to acknowledge discussions and contributions from  Sergey Bravyi, Antonio Corcoles, Andrew Cross, Easwar Magesan, Jim Rozen, Sarah Sheldon, and John Smolin. We acknowledge support from IARPA under contract W911NF-10-1-0324 and ARO under contract W911NF-14-1-0124.
\end{acknowledgments}



\clearpage

\end{document}